\RequirePackage{lineno}
\documentclass[twocolumn,showpacs,preprintnumbers,amsmath,superscriptaddress,amssymb]{revtex4}
\usepackage{graphicx}% Include figure files
\usepackage{lineno}% bold math
\begin{document}
%\setpagewiselinenumbers
%\modulolinenumbers[5]
%\linenumbers
%\preprint{APS/123-QED}

\title{A study of the associated production of photons and $b$-quark jets in $p \overline{p}$ collisions at $\sqrt{s}=1.96$ TeV}% Force line breaks with \\
%\maketitle
\affiliation{Institute of Physics, Academia Sinica, Taipei, Taiwan 11529, Republic of China} 
\affiliation{Argonne National Laboratory, Argonne, Illinois 60439} 
\affiliation{University of Athens, 157 71 Athens, Greece} 
\affiliation{Institut de Fisica d'Altes Energies, Universitat Autonoma de Barcelona, E-08193, Bellaterra (Barcelona), Spain} 
\affiliation{Baylor University, Waco, Texas  76798} 
\affiliation{Istituto Nazionale di Fisica Nucleare Bologna, $^{cc}$University of Bologna, I-40127 Bologna, Italy} 
\affiliation{Brandeis University, Waltham, Massachusetts 02254} 
\affiliation{University of California, Davis, Davis, California  95616} 
\affiliation{University of California, Los Angeles, Los Angeles, California  90024} 
\affiliation{University of California, San Diego, La Jolla, California  92093} 
\affiliation{University of California, Santa Barbara, Santa Barbara, California 93106} 
\affiliation{Instituto de Fisica de Cantabria, CSIC-University of Cantabria, 39005 Santander, Spain} 
\affiliation{Carnegie Mellon University, Pittsburgh, PA  15213} 
\affiliation{Enrico Fermi Institute, University of Chicago, Chicago, Illinois 60637}
\affiliation{Comenius University, 842 48 Bratislava, Slovakia; Institute of Experimental Physics, 040 01 Kosice, Slovakia} 
\affiliation{Joint Institute for Nuclear Research, RU-141980 Dubna, Russia} 
\affiliation{Duke University, Durham, North Carolina  27708} 
\affiliation{Fermi National Accelerator Laboratory, Batavia, Illinois 60510} 
\affiliation{University of Florida, Gainesville, Florida  32611} 
\affiliation{Laboratori Nazionali di Frascati, Istituto Nazionale di Fisica Nucleare, I-00044 Frascati, Italy} 
\affiliation{University of Geneva, CH-1211 Geneva 4, Switzerland} 
\affiliation{Glasgow University, Glasgow G12 8QQ, United Kingdom} 
\affiliation{Harvard University, Cambridge, Massachusetts 02138} 
\affiliation{Division of High Energy Physics, Department of Physics, University of Helsinki and Helsinki Institute of Physics, FIN-00014, Helsinki, Finland} 
\affiliation{University of Illinois, Urbana, Illinois 61801} 
\affiliation{The Johns Hopkins University, Baltimore, Maryland 21218} 
\affiliation{Institut f\"{u}r Experimentelle Kernphysik, Karlsruhe Institute of Technology, D-76131 Karlsruhe, Germany} 
\affiliation{Center for High Energy Physics: Kyungpook National University, Daegu 702-701, Korea; Seoul National University, Seoul 151-742, Korea; Sungkyunkwan University, Suwon 440-746, Korea; Korea Institute of Science and Technology Information, Daejeon 305-806, Korea; Chonnam National University, Gwangju 500-757, Korea; Chonbuk National University, Jeonju 561-756, Korea} 
\affiliation{Ernest Orlando Lawrence Berkeley National Laboratory, Berkeley, California 94720} 
\affiliation{University of Liverpool, Liverpool L69 7ZE, United Kingdom} 
\affiliation{University College London, London WC1E 6BT, United Kingdom} 
\affiliation{Centro de Investigaciones Energeticas Medioambientales y Tecnologicas, E-28040 Madrid, Spain} 
\affiliation{Massachusetts Institute of Technology, Cambridge, Massachusetts  02139} 
\affiliation{Institute of Particle Physics: McGill University, Montr\'{e}al, Qu\'{e}bec, Canada H3A~2T8; Simon Fraser University, Burnaby, British Columbia, Canada V5A~1S6; University of Toronto, Toronto, Ontario, Canada M5S~1A7; and TRIUMF, Vancouver, British Columbia, Canada V6T~2A3} 
\affiliation{University of Michigan, Ann Arbor, Michigan 48109} 
\affiliation{Michigan State University, East Lansing, Michigan  48824}
\affiliation{Institution for Theoretical and Experimental Physics, ITEP, Moscow 117259, Russia} 
\affiliation{University of New Mexico, Albuquerque, New Mexico 87131} 
\affiliation{Northwestern University, Evanston, Illinois  60208} 
\affiliation{The Ohio State University, Columbus, Ohio  43210} 
\affiliation{Okayama University, Okayama 700-8530, Japan} 
\affiliation{Osaka City University, Osaka 588, Japan} 
\affiliation{University of Oxford, Oxford OX1 3RH, United Kingdom} 
\affiliation{Istituto Nazionale di Fisica Nucleare, Sezione di Padova-Trento, $^{dd}$University of Padova, I-35131 Padova, Italy} 
\affiliation{LPNHE, Universite Pierre et Marie Curie/IN2P3-CNRS, UMR7585, Paris, F-75252 France} 
\affiliation{University of Pennsylvania, Philadelphia, Pennsylvania 19104}
\affiliation{Istituto Nazionale di Fisica Nucleare Pisa, $^{ee}$University of Pisa, $^{ff}$University of Siena and $^{gg}$Scuola Normale Superiore, I-56127 Pisa, Italy} 
\affiliation{University of Pittsburgh, Pittsburgh, Pennsylvania 15260} 
\affiliation{Purdue University, West Lafayette, Indiana 47907} 
\affiliation{University of Rochester, Rochester, New York 14627} 
\affiliation{The Rockefeller University, New York, New York 10021} 
\affiliation{Istituto Nazionale di Fisica Nucleare, Sezione di Roma 1, $^{hh}$Sapienza Universit\`{a} di Roma, I-00185 Roma, Italy} 

\affiliation{Rutgers University, Piscataway, New Jersey 08855} 
\affiliation{Texas A\&M University, College Station, Texas 77843} 
\affiliation{Istituto Nazionale di Fisica Nucleare Trieste/Udine, I-34100 Trieste, $^{ii}$University of Trieste/Udine, I-33100 Udine, Italy} 
\affiliation{University of Tsukuba, Tsukuba, Ibaraki 305, Japan} 
\affiliation{Tufts University, Medford, Massachusetts 02155} 
\affiliation{Waseda University, Tokyo 169, Japan} 
\affiliation{Wayne State University, Detroit, Michigan  48201} 
\affiliation{University of Wisconsin, Madison, Wisconsin 53706} 
\affiliation{Yale University, New Haven, Connecticut 06520} 
\author{T.~Aaltonen}
\affiliation{Division of High Energy Physics, Department of Physics, University of Helsinki and Helsinki Institute of Physics, FIN-00014, Helsinki, Finland}
\author{J.~Adelman}
\affiliation{Enrico Fermi Institute, University of Chicago, Chicago, Illinois 60637}
\author{B.~\'{A}lvarez~Gonz\'{a}lez$^v$}
\affiliation{Instituto de Fisica de Cantabria, CSIC-University of Cantabria, 39005 Santander, Spain}
\author{S.~Amerio$^{dd}$}
\affiliation{Istituto Nazionale di Fisica Nucleare, Sezione di Padova-Trento, $^{dd}$University of Padova, I-35131 Padova, Italy} 

\author{D.~Amidei}
\affiliation{University of Michigan, Ann Arbor, Michigan 48109}
\author{A.~Anastassov}
\affiliation{Northwestern University, Evanston, Illinois  60208}
\author{A.~Annovi}
\affiliation{Laboratori Nazionali di Frascati, Istituto Nazionale di Fisica Nucleare, I-00044 Frascati, Italy}
\author{J.~Antos}
\affiliation{Comenius University, 842 48 Bratislava, Slovakia; Institute of Experimental Physics, 040 01 Kosice, Slovakia}
\author{G.~Apollinari}
\affiliation{Fermi National Accelerator Laboratory, Batavia, Illinois 60510}
\author{A.~Apresyan}
\affiliation{Purdue University, West Lafayette, Indiana 47907}
% ...

\collaboration{CDF Collaboration\footnote{With visitors from $^a$University of Massachusetts Amherst, Amherst, Massachusetts 01003,
$^b$Universiteit Antwerpen, B-2610 Antwerp, Belgium, 
$^c$University of Bristol, Bristol BS8 1TL, United Kingdom,
$^d$Chinese Academy of Sciences, Beijing 100864, China, 
$^e$Istituto Nazionale di Fisica Nucleare, Sezione di Cagliari, 09042 Monserrato (Cagliari), Italy,
$^f$University of California Irvine, Irvine, CA  92697, 
$^g$University of California Santa Cruz, Santa Cruz, CA  95064, 
$^h$Cornell University, Ithaca, NY  14853, 
$^i$University of Cyprus, Nicosia CY-1678, Cyprus, 
$^j$University College Dublin, Dublin 4, Ireland,
$^k$University of Edinburgh, Edinburgh EH9 3JZ, United Kingdom, 
$^l$University of Fukui, Fukui City, Fukui Prefecture, Japan 910-0017
$^m$Kinki University, Higashi-Osaka City, Japan 577-8502
$^n$Universidad Iberoamericana, Mexico D.F., Mexico,
$^o$University of Iowa, Iowa City, IA  52242,
$^p$Kansas State University, Manhattan, KS 66506
$^q$Queen Mary, University of London, London, E1 4NS, England,
$^r$University of Manchester, Manchester M13 9PL, England,
$^s$Muons, Inc., Batavia, IL 60510, 
$^t$Nagasaki Institute of Applied Science, Nagasaki, Japan, 
$^u$University of Notre Dame, Notre Dame, IN 46556,
$^v$University de Oviedo, E-33007 Oviedo, Spain, 
$^w$Texas Tech University, Lubbock, TX  79609, 
$^x$IFIC(CSIC-Universitat de Valencia), 56071 Valencia, Spain,
$^y$Universidad Tecnica Federico Santa Maria, 110v Valparaiso, Chile,
$^z$University of Virginia, Charlottesville, VA  22906
$^{aa}$Bergische Universit\"at Wuppertal, 42097 Wuppertal, Germany,
$^{bb}$Yarmouk University, Irbid 211-63, Jordan
$^{jj}$On leave from J.~Stefan Institute, Ljubljana, Slovenia, 
}}
\noaffiliation

\begin{abstract}

The cross section for photon production in association with at least one jet containing a $b$-quark hadron
has been measured in proton antiproton collisions at 
$\sqrt{s}=1.96$ TeV.
The analysis uses a data sample corresponding to an integrated luminosity of 340 pb$^{-1}$ collected with the CDF II detector.
Both the differential cross section as a function of photon transverse energy
$E_T^{\gamma}$, $d \sigma$($p \overline{p} \rightarrow \gamma + \geq 1 b$-jet)/$d E_T^{\gamma}$
and the total cross section
$\sigma$($p \overline{p} \rightarrow \gamma + \geq 1 b$-jet; $E_T^{\gamma}> 20$ GeV) are measured.
Comparisons to a next-to-leading order prediction of the process are presented.

%Events have been selected using two trigger strategies: an isolated high
%transverse energy ($E_T^{\gamma}> 25$ GeV) photon, and a photon of lower transverse energy
%($E_T^{\gamma}>20$ GeV) produced in association with a jet and displaced track.
%
%A dataset of 340 pb$^{-1}$ has been used to perform the analysis. The inclusive
%cross section for photons of transverse energy exceeding 20 GeV, within pseudorapidities
%of $\pm 1.1$ produced in association with an identified $b$-quark hadron jet of transverse
%energy greater than 20 GeV, within pseudorapidities of $\pm 1.5$, is $54.82 \pm 3.3 \pm 5.1$ pb,
%where the first error is statistical and the second systematic. 

%The inclusive cross sections obtained is:
%
%\[\sigma(b+\gamma, |\eta(b)|<1.5, |\eta(\gamma)|<1.1, \Delta R(b-\gamma)>0.7,
%Et(b) > 20\]
%\[ E_T(\gamma)>26) = 34.8 \pm 3.4 (stat)^{+7.2}_{-7.4}(syst.) pb \]

%\[\sigma(b+\gamma, |\eta(b)|<1.5, |\eta(\gamma)|<1.1, \Delta R(b-\gamma)>0.7,
%Et(b) > 20\]
%\[ E_T(\gamma)>20) = 54.2 \pm 3.3 (stat) \pm 5.1 (syst.) pb \]

%This result is in agreement with next-to-leading order QCD predictions.

\end{abstract}

\pacs{Valid PACS appear here}% PACS, the Physics and Astronomy
                             % Classification Scheme.
%Use showkeys class option if keyword
                              %display desired
\maketitle

%\section{\label{sec:level1}First-level heading:\protect\\ The line
%break was forced \lowercase{via} \textbackslash\textbackslash}

% 
%  introduction
%
%\section{Introduction}                                               

The study of proton-antiproton interactions with an isolated, 
high energy photon and an 
identified $b$-quark jet is a testing ground for 
quantum chromodynamics (QCD) predictions 
at the Tevatron.
At photon transverse energies $E_T^{\gamma}$ below 70 GeV Compton scattering processes dominate production,
with $gb \rightarrow \gamma gb$ or $qb \rightarrow \gamma q b$ dominating at lower $E_T^{\gamma}$
depending on the relative sizes of the quark and gluon parton density functions. Above an $E_T^{\gamma}$ of 70 GeV
$q \overline{q} \rightarrow b \overline{b} \gamma$
quark annhiliation processes dominate production \cite{nlo}.
A cross section measurement therefore provides a
well measured probe of the hard scattering dynamics 
within the proton, and some sensitivity
to the $b$-quark content of the proton,
whose parton density function is so far indirectly extracted from constraints on the gluon density 
functions.

%%
%  not sure if we need this now?
%
%Recent measurements of jets containing a $b$-quark hadron (named $b$-jet in the following), in conjunction with
%a Z boson~\cite{zbnew} have shown a
%small excess in data, compared
%to leading and next-to-leading order predictions calculated at different scales.
%A measurement of the photon and $b$-jet production cross section
%%which should be measureable with higher statistics,
%may therefore yield deeper insight into the underlying QCD mechanisms
%of $b$-jet production in association with a boson.

%It is important that this process is well understood, as the experimental
%signature of a photon and b hadron jet
%can also be replicated by many models of new physics. Most hypotheses
%of new physics involve the production of massive objects which can often 
%decay to heavy quarks. Photons may be produced by radiative decays of 
%heavy states (eg. Techni-Omega), or in decays of 
%supersymmetric particles in gauge-mediated models.
%An example of the latter is 
%the production of chargino-neutralino pairs and subsequent decay 
%through intermediate stop production,
%$\chi_1^+ \chi_2^0 \rightarrow bc \chi_1^0 \gamma \chi_1^0$, a mechanism  
%invoked as an 
%explanation of an anomalous event containing two electrons, two photons and 
%missing energy observed 
%in an earlier data-taking period at the Tevatron~\cite{eggmet}.

The first measurement of photon and heavy flavour jet production was performed using 
86 pb$^{-1}$ of integrated luminosity taken at $\sqrt{s}=1.8$ TeV with the CDF 
I detector \cite{cdfrun1}. Heavy jet flavor was signalled by muons contained in the jet. The
results were interpreted in terms of new physics processes involving the radiative
decay of Techni-Omega states~\cite{techniomega}, and radiative decays of
supersymmetric particles in gauge-mediated supersymmetry models~\cite{cdfphob}.
Recently the D0 collaboration has measured the cross section of heavy flavour jet production in
association with a photon~\cite{d0pub} using data collected at $\sqrt{s}=1.96$ TeV.
In this paper we 
exploit improvements in the CDF II detector to identify b jets by a lifetime based
secondary vertex
tag,
use a larger dataset collected at a higher center-of-mass
energy, probe lower photon transverse energies,  
and employ a superior analysis technique where all backgrounds are
determined directly from data. 
We present our results as 
the differential photon + b jet production cross section, as a function of photon $E_T^{\gamma}$,
and the total photon + b jet production cross section.

%
% detector
%
%\section{Detector}

The CDF II detector is described in detail in \cite{cdfii}. 
Only the components which are most relevant to this analysis will
be described here.

The detector is composed of a central spectrometer inside a 1.4T magnetic field,
surrounded by electromagnetic and hadronic calorimetry and muon
chambers. 
The spectrometer is composed of a multi-layer silicon vertex detector
inside a cylindrical multi-wire drift chamber.
The combination of these tracking
detectors measures charged particle trajectories with a transverse momentum ($p_T$)
precision of $\Delta p_T/p_T^2 = 0.07\%$ (GeV/c)$^{-1}$, and an uncertainty on 
the transverse impact parameter
of about 40 $\mu$m for tracks of $p_T$ above 1 GeV/c, 
which includes the intrinisic beam size of about
30 $\mu$m. Information from the central tracker can be sent within trigger latency
to the hardware tracker SVT\cite{svt}, that provides an extra-fast measurement of the tracking 
parameters the can be used at trigger level.

Segmented sampling calorimeters, arranged in a projective tower geometry, surround the tracking detectors and provide energy
measurements within the region $\mid\eta\mid<3.6$.
Central calorimeters~\cite{central} cover the region $\mid\eta\mid<1.1$, with an electromagnetic (hadronic)
energy resolution of
$\sigma(E)/E = 13.5 \%/ \sqrt{E} \oplus 2.0 \%$ 
($\sigma(E)/E = 50\%/\sqrt{E_T} \oplus 3\%$).
The end-wall hadronic calorimeter extends this coverage to $\mid\eta\mid<1.3$~\cite{wall}
with an energy  resolution of $75\%/\sqrt{E_T} \oplus 4\%$,
whilst the region $1.3<\mid\eta\mid<3.6$ ($1.1<\mid\eta\mid<3.6$ for the electromagnetic
calorimeter) is covered by forward calorimeters~\cite{plug}, with hadronic and electromagnetic energy resolutions
of
$80\%/\sqrt{E} \oplus 5\%$
and
$16\%/\sqrt{E} \oplus 1\%$ respectively. 
%
%The energy
%%resolution of the electromagnetic (hadronic) calorimeter within the region $\mid\eta\mid<1.1
%%is $\sigma(E)/E = 13.5 \%/ \sqrt{E} \oplus 2.0 \%$~\cite{cem} 
%%($\sigma(E)/E = 50\%/\sqrt{E} \oplus 3\%$~\cite{cha}). In the region $1.1 < \mid\eta\mid < 1.5$
%%the resolution is $\sigma(E)/E = 14.4\% / \sqrt{E} \oplus 0.7\%$~\cite{plug}
%%($ $~\cite{wha}).%
%
In order to distinguish electromagnetic clusters from
photons and electrons, and clusters from neutral pion decays, the
central electromagnetic calorimeter is equipped with two sets of wire chambers:
a preshower detector (CPR) situated in front of the active part of the calorimeter
to detect early photon conversions in the solenoid coil, and a
shower maximum detector (CES) placed at the position of maximal width of 
electromagnetic showers to measure the shower profile. 
Signals from these detectors are compared with
expected deposits from electromagnetic clusters, obtained from testbeam data.
Each electromagnetic cluster is given a weight related
to the probability that it originates from a photon.

%Particularly
%relevant to this analysis is the presence of the silicon vertex tracker (SVT)
%\cite{svt} in the trigger. 
%This device compares hits from the tracking detectors with 
%pre-fitted tracks stored in an associative memory to extract track parameters 
%online. In this way tracks with significant impact parameters can be quickly 
%identified. The impact parameter resolution of this procedure is 47 $\mu$m, 
%similar to the accuracy obtained
%offline. 
%, being very useful to provide datasets enriched in heavy-flavor
%mesons or jets.
\par

%
% datasets and trigger
%
%\section{Datasets and trigger}

We use data obtained by two triggers; a trigger which requires
a photon-like object with transverse energy
larger than the threshold for the inclusive photon trigger at CDF of 25 GeV 
(`high $E_T^{\gamma}$ photon'), and a trigger (`SVT photon') which 
requires a 
photon-like object with transverse energy larger than 12 GeV, a jet with
transverse energy larger than 10 GeV, and a track, measured by the SVT~\cite{svt},
with transverse momentum larger than 2 GeV/c, an absolute rapidity smaller than
1 and an impact parameter larger than 120 $\mu$m. 
The SVT compares hits from the tracking detectors with 
pre-fitted tracks stored in an associative memory to extract track parameters 
in real time. In this way tracks with significant impact parameters can be quickly 
identified. The impact parameter resolution of this procedure is less than
50 $\mu$m (including the contribution from the beam size), 
similar to the accuracy obtained
with full detector reconstruction. 
An integrated luminosity of 340 (208) pb$^{-1}$ of data has been analyzed in
the high $E_T^{\gamma}$ photon (SVT photon) triggered dataset.
%Both triggers
%are unprescaled for the data taking period considered here, and the larger
%phase space accessible with the dedicated trigger comes at the price of 
%efficiency loss due to the additional tracking requirement.

The two triggers have very different efficiencies.
The high $E_T^{\gamma}$ photon trigger
has an efficiency close to 100\% for events with $E_T^{\gamma}$ above 28 GeV. 
To obtain photons of lower $E_T^{\gamma}$ the SVT trigger must be used.
The efficiency of the
SVT photon trigger is found by comparing the number of events
that pass both triggers. Being based on strong requirements on a single track
rather than a more refined secondary vertex reconstruction, the efficiency of 
the SVT-based photon trigger is about 50\% (with a relative uncertainty of $4\%$, 
due to the size of the overlap sample) and was found to be independent of 
the jet transverse energy. 

%\section{Selection}

Selected candidate events must pass one of the two photon triggers, contain an isolated 
central ($\mid\eta\mid < 1.1$) 
photon of $E_T^{\gamma}> 20$ GeV, and a b jet 
of $E_T > 20$ GeV within $\mid \eta \mid < 1.5$. 

Photon candidates must have a calorimeter cluster which contains predominately electromagnetic energy. 
The cluster 
hadronic energy fraction $E_{HAD}/E_{EM}$ must satisfy $E_{HAD}/E_{EM} <
0.055 + 0.00045 * E^{\gamma}$, where $E^{\gamma}$ is the photon energy and $E_{HAD}$ and $E_{EM}$ are the hadronic and electromagnetic
energy deposits within the cluster. The electromagnetic shower profile
must also agree with that expected for an electromagnetic deposit.
In order to reduce contamination from neutral meson (such as $\pi^0$) decays, photon
candidates must be isolated from nearby calorimetric deposits and tracks.
We require that $E_T(R^{0.4}) < 2.0 + 0.02 (E_T^{\gamma}-20)$, where $E_T(R^{0.4})$ is the summed
calorimetric cluster transverse energy deposits in a cone of radius 
$R=\sqrt{\Delta \phi^2 + \Delta \eta^2} =$
0.4 around the photon candidate, to ensure
calorimetric isolation. In addition, the summed track transverse momenta of tracks inside this cone, $p_T(R^{0.4})$ must 
satisfy $p_T(R^{0.4}) < 2.0 + E_T^{\gamma}*0.005$ to ensure isolation in the tracking detectors.
Events containing adjacent calorimetric clusters in the CES, which can signal the presence of neutral mesons, are rejected.

%Photon candidates are identified by placing requirements on:
%the ratio of hadronic to electromagnetic energy deposited, the agreement of the 
%electromagnetic shower profile with that expected for an electromagnetic deposit,
%calorimetric energy and summed track transverse momenta in a cone around the photon candidate,
%and the adjacent calorimetric clusters to ensure isolation.

The jets in each event are reconstructed using the JETCLU algorithm~\cite{jetclu}
with cone radius R = 
0.4 (0.7) for events containing photons of $E_T^{\gamma}< (>) 26$ GeV. 
To recover the true parton energy, jets are corrected for 
instrumental effects \cite{jetcor}.
We select events containing at least one jet with corrected
$E_T > 20$ GeV, and whose
axis lies outside a 
cone of $\Delta R = 0.7$ surrounding the photon candidate.
In order to identify jets arising from b hadrons we search for displaced
secondary vertices \cite{secvtx}. 
A b jet is 
identified (`tagged') when the secondary vertex is more than two 
standard deviations away from the beam position, and in the same direction 
away from the beam position as the jet momentum. We require at least one
b jet to be identified in each event. The efficiency of the $b$-tagging
algorithm is just under 25\% for b jets of $E_T=20$ GeV, and increases to 40$\%$ at $E_T=50$ GeV.

The {\sc pythia}~\cite{pythia} Monte Carlo code
is used to simulate photon + jet production at leading order and to estimate the photon and jet selection
efficiencies. The 
$Q^2$ scale of the interaction is set to 225 GeV$^{2}$, and the
CTEQ5L~\cite{cteq} parton distribution functions are used.
A simulation of the underlying event is included~\cite{tunea}.

%
% selection and backgrounds
%

\begin{figure}

\includegraphics[angle=270,width=7cm]{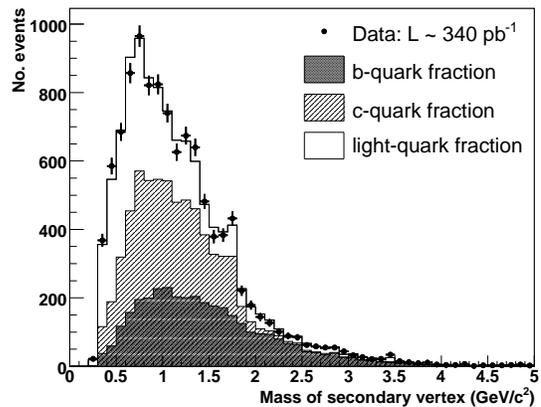}

\caption{Fit to the invariant mass of tracks composing the secondary vertex in
data, for photon candidates having $E_T > 26$ GeV. The points are data, and the stacked, shaded 
histograms represent the estimated contributions of the b, c- and light-quark 
jets.}\label{fit}
\end{figure}

Backgrounds to photons arise from high energy $\pi^0$'s, which decay to pairs of 
overlapping
photons that cannot be distinguished in the calorimeter, but whose rate can be estimated directly
from the data.
The fraction of misidentified photons is estimated using
the
clusters detected by the CPR and CES detectors of the 
electromagnetic calorimeter. 
Photon pairs produced by $\pi^0$ decays will produce
a larger number of hits in these detectors, and each photon candidate
is assigned a weight
proportional to the probability of being a signal or background cluster following the procedure detailed in~\cite{cescpr}. 
The fraction of correctly identified photons in the sample passing those selection 
criteria increases with increasing $E_T$, going from about 50\% at the lower end of 
the spectrum considered here (20 GeV) to around 80\% at high $E_T^\gamma$.
Backgrounds to b jets can arise from jets originating from charm quarks, since charm
hadrons have a lifetime
between a quarter and two thirds that of $b$-hadrons, and jets originating from light quarks and gluons, where 
random combinations of tracks can sometimes mimic a
vertex displaced from the beamline.
The purity of our selected events is determined by fitting the invariant mass
of tracks composing the secondary vertex in data
to Monte Carlo templates of the shapes 
expected for b-, charm (c-) and light-quark jets. 
Fig. ~\ref{fit} shows an example of the fit to data. Here, about a third
of jets arise from b quarks. 
This invariant mass is generally lower than
the corresponding hadron mass due to mis-assigned tracks and unreconstructed
neutral hadrons. 
It can be seen that the template shapes of the different quark jet types are 
sufficiently different to provide reasonable discriminating power.

%\begin{figure}
%\includegraphics[angle=270,width=7cm]{templates.epsi}
%\caption{Secondary vertex invariant mass distributions for b, c and light jets.}\label{templates}
%\end{figure}

Identified b jets can arise from either the photon + b jet signal,
or 
from events where one or both objects (photon and jet) are mis-identified.
We assume that the tagged jet composition in $\pi^0$ + tagged jet events is similar
to $\pi^{\pm}$ + tagged jet events.
We estimate the b jet purity of the misidentified
photon background using data collected with a jet based trigger. Events are
required to contain two jets, one of which must be tagged and have similar transverse energy
and pseudorapidity requirements to the b jet, and a second
which passes similar kinematic requirements to the photon. The fraction
of b jets in this
sample can be found by fitting the secondary vertex invariant mass of the 
tagged jets.
The purity of the selected jets ranges from 50\% for jets of $E_T$ 
around 20 GeV, to about 15\% for
jets of $E_T$ around 75 GeV
where the rate of light-quark jet tagging increases.
This b fraction is then normalised to the estimated number of misidentified
photons, and subtracted from the estimated number of b jets in the 
whole event sample to yield the total number of photon + b jet events.
\par

Some 10900 (55800) events pass the selection criteria in the 
high transverse energy photon
(SVT photon) triggered datasets.
Candidate events are divided into bins of photon transverse energy. The numbers of 
events in each bin are corrected for background contributions, 
trigger, selection and acceptance 
efficiency, and divided by the appropriate integrated luminosity to yield a
cross section. Results are given in Table~\ref{results}, which also lists
the systematic uncertainties detailed later. 
Note that the statistical uncertainty for the high $E_T^{\gamma}$
photon dataset includes contributions from finite Monte Carlo statistics.

%/includegraphics[width=7cm]{../mor07/trigefficiency.eps}% Here is how to import EPS art
%\caption{\label{fig:trigeff}Efficiency of the SVT-based trigger as a function of
%jet Et}

%\end{figure}

%
%  results
%

%The total cross section has been measured first in the high photon Et region on the 
%unbiased dataset. The integrated cross section, measured on
%an integrated luminosity of  340 pb$^{-1}$, is 
%\[\sigma(b+\gamma, |\eta(b)|<1.5, |\eta(\gamma)|<1.1, \Delta R(b-\gamma)>0.7,
%Et(b) > 20\]
%\[ Et(\gamma)>26) = 42.0 \pm 3.8 pb \]
%This result is used to calculate the efficiency of the SVT-based trigger, and extend
%the measurement to the SVT-based dataset in the low photon Et region. For reasons of 
%trigger stability, this
%analysis is restricted to 208 pb$^{-1}$ of data, and the result is
%\[\sigma(b+\gamma, |\eta(b)|<1.5, |\eta(\gamma)|<1.1, \Delta R(b-\gamma)>0.7,
%Et(b) > 20\]
%\[ Et(\gamma)>12) = 90.5 \pm 6.0 pb \]
%to be compared to a leading-order prediction from pythia of 69.3 pb. The differential
%cross section as a function of photon Et is shown in Fig. \ref{fig:photet12}.

%
% systematics
%
%\section{Systematics}
%\label{sys}

Several sources of systematic uncertainty
in the cross section determination have been studied:
photon identification, jet energy scale, b jet identification, and
luminosity. In the following only the largest contributions
will be quantified.

Variables used in photon identification have been studied in
$Z \rightarrow e^{+} e^{-}$ simulation and data
events ~\cite{diphoton}. Simulation and data are consistent.
Uncertainties in the misidentified photon estimate arise from
assumed values for the preshower detector hit rate, the rate of backscattered showers 
and the fractional composition of fake photon backgrounds. The associated
systematic uncertainty is about 6\%.

The systematic uncertainty on the jet energy scale has been studied in detail 
elsewhere~\cite{jetcor} and the findings applied to this analysis. It decreases
with increasing jet $E_T$,
and is about $5\%$ for jets of 35 GeV $E_T$.
Uncertainties
have also been determined for the effect of multiple interactions overlapping in the same data event,
and the additional uncertainty on the jet scale introduced by a b jet.

Uncertainties in the b quark purity of the sample arise from imperfect modelling of the 
template shapes used in fits to extract the b purity. Differences in shape 
can arise if the secondary vertex invariant mass of jets containing one (`single')
or two (`double') b quarks
differs, or if track efficiency is incorrectly modelled.
We find that template shapes 
obtained from single and double quark jets 
are not consistent with each other. The data are fitted to templates composed of mixtures of 
the individual single
and double quark templates (ranging from 0 to 100\%), and the $\chi^2$ 
of the resulting distributions with respect to the data is computed. 
We take as a $1-\sigma$ deviation the value for this mixture for which the
$\chi^2$ increases by one unit with respect to its minimum value, and recalculate
the cross section using this mixed template. The systematic uncertainty is the difference
between this value and the default cross section (obtained with the default diquark
fraction).

The uncertainty on the cross section is assigned as the difference between the 
reference value and that obtained from the value of the single/double quark template 
for which the $\chi^2$
increases by one unit with respect to its minimum value.
This is the largest single source of uncertainty and is about 17\%.
Previous studies ~\cite{zb} suggest a difference in tracking efficiency
between data and simulation which is a function of isolation,
momentum, and position in the detector. We remake the invariant mass templates 
incorporating the inefficiency derived from data, and take the full difference 
between results obtained using these
and the nominal templates (5\%) as the systematic uncertainty due to this source. 

Other systematic uncertainties on b jet identification 
arise from the difference in measured
tagging efficiency between data and simulation, from
the difference in the tagging efficiency of single and double b jets, 
and from simulated b hadron multiplicity. 
The first uncertainty has been previously studied~\cite{secvtx}.
The difference in scale between tagging efficiency in data and simulation
was found to be $0.91 \pm 0.06$. This results in a $6\%$ uncertainty on the measured cross-sections.
The uncertainty on tagging efficiency for single and double b jets is determined as 
the difference between results obtained using 
the fractions of single and double quark templates corresponding to $\pm 1$ standard 
deviation, as found earlier, and is about 7\%.
We have adopted the findings of previous
studies of the effect of assumed b hadron multiplicity~\cite{zb} (a $1\%$ effect on the measured cross section).
The SVT-based analysis is also affected by the statistical precision 
of the trigger efficiency determination (about 10\%).
Finally, the luminosity is subject to a $\pm 6\%$ uncertainty \cite{lumi}.

%All systematic errors are summarised in table \ref{tab:systall}.\par

%\begin{table}[tbh]
%\begin{center}
%\begin{tabular}{|l|c|c|c|c|} \hline

%Photon Et (GeV) & Photon identification & b jet identification & 
%luminosity & Total \\ 
%/ Source & & & & \\ \hline
%%%%26-28 &  $ \pm 0.19$ &  $\pm 0.61$ & $\pm 0.19$ & $\pm 0.67$ \\
%28-31 &   $ ^{+ 0.15}_{ - 0.18}$ & $ ^{+ 0.57}_{ - 0.56}$ & $\pm 0.17$ & $\pm 0.61$ \\
%31-35 & $\pm 0.1$ & $\pm 0.24$ & $ \pm 0.07$ & $\pm 0.27$ \\
%35-43 & $ ^{+0.03}_{ - 0.06}$ & $^{+0.17}_{ - 0.18}$ & $\pm 0.06$ & $^{+0.18}_{ - %0.20}$ \\
%43-70 & $ ^{+0.010}_{ - 0.016}$ & $^{+0.037}_{ - 0.039}$ & $\pm 0.012$ & 
%$^{+0.040}_{ - 0.044}$ \\ \hline

%\end{tabular}
%\caption{Systematic errors for each bin of photon Et, for the sources
%discussed in the text: photon identification; b jet identification;
%luminosity. All errors are given 
%as a percentage.\label{tab:systall}}
%\end{center}
%\end{table}

%%\begin{center}
%\begin{tabular}{|l|c|} \hline
%Source&Syst. error (\%)\\ \hline
%Luminosity&6\\
%Photon ID&1\\
%Photon BG&$^{+4}_{-6}$\\
%Energy scale&4\\
%B purity&10\\
%$b\bar{b}$ fraction&2\\
%Tagging eff.&3\\ \hline
%SVT eff.&10\\ \hline
%\end{tabular}
%\caption{Summary of systematic errors\label{tab:syst}}
%\end{center}
%\end{table}

\par

%\section{Results}

The cross section for photons produced in association with b jets
is tabulated in Table~\ref{results}.
% and shown in Figure~\ref{fig:photet12}. 
Both differential and inclusive results
are given, with their statistical and systematic uncertainties. 
Note that results from
both datasets are listed. There is overlap between the final SVT photon and the
first high $E_T^{\gamma}$ photon bins, which give consistent results. Due to its greater statistical precision, the former
bin is used in the final results.
The measurements are corrected to hadron level~\cite{hadron}, so that they can be directly compared
%to both the leading order hadron level result from {\sc pythia} and 
to a recent 
next-to-leading order calculation\cite{nlo}. 
This prediction was derived analytically, using the CTEQ6.6M parton density functions~\cite{cteq66}, and a 
renormalisation, factorisation and fragmentation scale set to the transverse 
momentum of the photon. 
It does not include non-perturbative effects (hadronisation and underlying event), and is presented in terms of parton level jets.

\begin{figure}
\includegraphics[width=8cm]{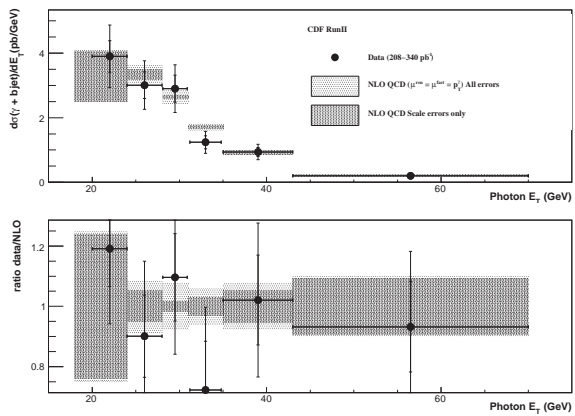}% Here is how to import EPS art
\caption{\label{fig:photet12} Top: b + photon cross section as a 
function of photon
$E_T$, compared to NLO QCD calculations. The light dashing is a quadrature sum
of uncertainties coming from scale variation (both renormalisation and factorisation
scales varied by a factor 2 and 0.5) and PDFs, while the darker dashing 
represents the scale variation contribution only. The inner error bars for data
represent the statistical uncertaintiess, the outer the combination of statistical
and sytematic. The bottom plot shows the ratio of data to NLO calculation, where the 
error bars and shading have the same meaning as before.}
\end{figure}

%NLO theory predicts a 30\% higher cross
%section, in better agreement with the experimental data.

\begin{table*}[tbh]
\begin{center}
\begin{tabular}{c c c} \hline
$E_T^{\gamma}$ (GeV) & $d \sigma$($p \overline{p} \rightarrow \gamma + \geq 1 b$-jet)/$d E_T^{\gamma}$(Data) & 
$d \sigma$($p \overline{p} \rightarrow \gamma + \geq 1 b$-jet)/$d E_T^{\gamma}$(NLO) \\ \hline \hline
 
20 - 24 & $3.90 \pm 0.49 \pm 0.84$ & $3.27 \pm 0.78$\\
24 - 28 & $3.01 \pm 0.41 \pm 0.63$ & $3.67 \pm 0.32$\\
26 - 28(*) & $3.13 \pm 0.51 \pm 0.67$ & $3.01 \pm 0.21$ \\
28 - 31 & $2.90 \pm 0.42 \pm 0.61$ & $2.65 \pm 0.18$ \\
31 - 35 & $1.24 \pm 0.20 \pm 0.27$ & $1.72 \pm 0.14$ \\
35 - 43 & $0.94 \pm 0.14 ^{+0.18}_{-0.20}$ & $0.92 \pm 0.10$ \\
43 - 70 & $0.20 \pm 0.03 \pm 0.04$  & $0.21 \pm 0.05$\\ \hline

%$>20$ & $54.22 \pm 3.26 ^{+5.04}_{-5.09}$ & $55.62 \pm 3.87$ \\
%\hline

\end{tabular}
\caption{The measured differential cross section for central photon production
in association with at least one $b$-jet of $E_T>20$ GeV, inside $\mid\eta\mid<1.5$, tabulated
as a function of the photon transverse energy $E_T^{\gamma}$. 
The first (second) uncertainty quoted is statistical (systematic).
Results are compared to a
next-to-leading order (NLO) prediction, whose uncertainty arises from the numerical integration procedure~\cite{nlo}. 
Note that the first two measurements are made using the SVT dataset,
and the remainder with the high $E_T^{\gamma}$ dataset. The overlap bin, denoted by (*), is not used in the final results. 
Systematics about luminosity, secondary vertex tagging scale factor, the effect of multiple
verteces are correlated between all bins. The uncertainty due to the statistical precision of
the SVT trigger efficiency is fully correlated between the first two bins only.
All other uncertainties are uncorrelated between bins.
\label{results}}
\end{center}
\end{table*}

The measured cross sections are shown compared with 
this prediction in Fig. ~\ref{fig:photet12}. Also shown are the theoretical uncertainties due to (i) choice of scale
and (ii) uncertainty in parton density function assumption. 
Agreement with next-to-leading order theory is good over the entire photon 
$E_T^{\gamma}$ range probed. It should be noted that due to numerical stability 
problems, the first bin in the NLO calculation starts at 18 GeV instead of 20 as
for the data. 

The total cross section $\sigma$($p \overline{p} \rightarrow \gamma + \geq 1 b$-jet; $E_T^{\gamma}> 20$ GeV)
has been measured to be $54.22 \pm 3.26$ (stat) $^{+5.04}_{-5.09}$ (syst) pb. This is consistent with the next-to-leading
order prediction of $55.62 \pm 3.87$ pb.

%
% conclusions
%
%\section{Conclusions}

In conclusion, the cross section for photon production in association with b jets
has been measured in proton antiproton collisions 
at $\sqrt{s}=1.96$ TeV with the CDF II detector.
The measurement has been made for $b$-jets with $E_T>20$ GeV inside $\mid\eta\mid<1.5$,
and for photons of at least $E_T^{\gamma}>20$ GeV inside $\mid\eta\mid<1.1$, including the lowest
photon transverse energies probed to date.
The results are consistent with next-to-leading order theoretical peturbative QCD predictions, using
CTEQ6.6M parton density functions, throughout the photon $E_T^{\gamma}$ range measured,
while leading-order calculations would predict a cross section smaller by about 30\%.
The level of accuracy of this measurement is therefore already sufficient to discriminate 
between the first orders of perturbative expansion, and favor the most precise NLO 
predictions.

%The differential cross section
%$d \sigma$($p \overline{p} \rightarrow \gamma + \geq 1 b$-jet)/$d E_T^{\gamma}$
%has been measured as a function of photon transverse energy, and the total 
%cross section 
%$\sigma$($p \overline{p} \rightarrow \gamma + \geq 1 b$-jet) has been measured
%for photons with $E_T>20$ GeV.  

%Events have been selected using two trigger strategies: requiring an isolated high
%transverse energy ($E_T > 25$ GeV) photon, and a photon of lower transverse energy
%($E_T >12$ GeV) produced in association with a jet and displaced track.
%The inclusive cross sections obtained is:
%
%$\sigma(b+\gamma, |\eta(b)|<1.5, |\eta(\gamma)|<1.1, \Delta R(b-\gamma)>0.7, Et(b) > 20  E_T(\gamma)>26) = 34.8 \pm 3.4 (stat)^{+7.2}_{-7.4}(syst.)$ pb;
%
%$\sigma(b+\gamma, |\eta(b)|<1.5, |\eta(\gamma)|<1.1, \Delta R(b-\gamma)>0.7, Et(b) > 20 E_T(\gamma)>20) = 54.2 \pm 3.3 \pm 5.1 (syst.)$ pb
%
%agree in the subset of events overlapping both triggers.
%These measurements are consistent with a next-to-leading order prediction, and about 30\% higher 
%than the LO predictions from {\sc pythia}.

We thank the Fermilab staff and the technical staffs of the participating institutions for their vital contributions. This work was supported by the U.S. Department of Energy and National Science Foundation; the Italian Istituto Nazionale di Fisica Nucleare; the Ministry of Education, Culture, Sports, Science and Technology of Japan; the Natural Sciences and Engineering Research Council of Canada; the National Science Council of the Republic of China; the Swiss National Science Foundation; the A.P. Sloan Foundation; the Bundesministerium f\"ur Bildung und Forschung, Germany; the World Class University Program, the National Research Foundation of Korea; the Science and Technology Facilities Council and the Royal Society, UK; the Institut National de Physique Nucleaire et Physique des Particules/CNRS; the Russian Foundation for Basic Research; the Ministerio de Ciencia e Innovaci\'{o}n, and Programa Consolider-Ingenio 2010, Spain; the Slovak R\&D Agency; and the Academy of Finland.


\begin{thebibliography}{99}

\bibitem{nlo} T.Stavreva and J.Owens, Phys. Rev. D {\bf 79}, 054017 (2009), and private 
communication relating predictions to the kinematic requirements used in the 
analysis presented here.
\bibitem{cdfrun1} T. Affolder et al. (CDF Collaboration), Phys. Rev. D {\bf 65}, 052006 (2002)
%\bibitem{zbnew}CDF Collaboration, hep-ex/0812.4458 (Submitted to Phys. Rev. D).
%
%\bibitem{eggmet} S.~Ambrosanio {\it et al.}, Phys. Rev. Lett. {\bf 76}, 3498 (1996) and Phys. Rev. D {\bf 55}, 1372 (1997).
\bibitem{techniomega} F.~Abe {\it et al.} (CDF Collaboration), Phys. Rev. D {\bf 60}, 092003 (1999).
\bibitem{cdfphob}T.~Affolder {\it et al.} (CDF Collaboration), Phys. Rev. D {\bf 65}, 052006 (2002).
\bibitem{d0pub} V.~M.~Abazov {\it et al.} (D0 Collaboration), Phys. Rev. Lett. {\bf 102}, 192002 (2009).
\bibitem{cdfii}  CDF Collaboration, FERMILAB-PUB-96/390-E, 1996, and Phys.\ Rev.\ D {\bf 68}, 072004 (2003).
%\bibitem{coord}In the CDF coordinate system, $\theta$ is the polar angle with 
%respect to the proton beam axis (positive z direction),
%and $\phi$ is the azimuthal angle. The pseudorapidity $\eta$ is defined by
%$\eta = -\ln( \tan(\theta/2) $. The transverse energy $E_T$ is the component
%of energy projected into the x-y plane. The impact parameter is defined to be
%the distance of closest approach of a track to the primary vertex.
\bibitem{svt}  W.~Ashmanskas {\it et al.}, Nucl. Instrum. Methods A {\bf 447}, 218 (2000).
\bibitem{central} L.~Balka {\it et al.}, Nucl. Instrum. Methods A {\bf 267}, 272 (1988).
\bibitem{wall} S.~Bertolucci {\it et al.},  Nucl. Instrum. Methods A {\bf 267}, 301 (1988).
\bibitem{plug} R.~Oishi, Nucl. Instrum. Methods A {\bf 453}, 277 (2000); \\
M.~G.~Albrow {\it et al.}, Nucl. Instrum. Methods A {\bf 480}, 524 (2002).
\bibitem{jetclu} F.~Abe {\it et al.} (CDF Collaboration), Phys. Rev. D {\bf 45}, 1448 (1992).
\bibitem{jetcor} A.~Bhatti {\it et al.}, Nucl. Instrum. Methods A {\bf 566}, 2 (2006).
\bibitem{secvtx} D.~Acosta {\it et al.} (CDF Collaboration), Phys. Rev. D {\bf 71}, 052003 (2005).
\bibitem{pythia} T.~Sjostrand {\it et al.}, JHEP 05, 026 (2006).
\bibitem{cteq} H.~L.~Lai {\it et al.}, Eur. Phys. J. C{\bf 12}, 375 (2000).
\bibitem{tunea} R.~Field, AIP Conf. Proc. 828:163 (2006).
\bibitem{cescpr} F.~Abe {\it et al.} (CDF Collaboration), Phys. Rev. D {\bf 48}, 2998 (1993).
\bibitem{diphoton} D.~Acosta {\it et al.} (CDF Collaboration), Phys. Rev. Lett. {\bf 95}, 022003 (2005).
\bibitem{zb} A.~Abulencia {\it et al.} (CDF Collaboration), Phys. Rev. D{\bf 74}, 032008 (2006).
\bibitem{lumi}S.~Klimenko, J.~Konigsberg, T.~M.~Liss, {\it FERMILAB-FN-0741} (2003).
\bibitem{hadron} The hadron level in Monte Carlo generators is defined using
all final state particles with lifetime above 10$^{-11}$ s.
\bibitem{cteq66} P.~M.~Nadolsky, Phys. Rev. D {\bf 78}, 013004 (2008).
\end{thebibliography}
\end{document}